\documentstyle[sprocl]{article}

\input{psfig}

\def\Journal#1#2#3#4{{#1} {\bf #2}, #3 (#4)}

\def\NPB{{\em Nucl.\ Phys.} B}
\def\PLB{{\em Phys.\ Lett.}  B}
\def\PRL{\em Phys.\ Rev.\ Lett.}
\def\PRD{{\em Phys.\ Rev.} D}

\def\be{\begin{equation}}
\def\ee{\end{equation}}
\def\bea{\begin{eqnarray}}
\def\eea{\end{eqnarray}}

\begin{document}

\title{GLUEBALLS FROM IMPROVED LATTICE ACTIONS}

\author{COLIN MORNINGSTAR$^{a}$ and MIKE PEARDON$^{b}$ }
\address{ ${}^a$University of California at San Diego, La Jolla,
 CA 92093-0319 \\
 ${}^b$University of Kentucky,  Lexington,
 KY 40506-0055 }

\maketitle\abstracts{
The low-lying glueball masses and the hadronic scale $r_0$ are
computed in lattice SU(3) gauge theory with the aim of establishing
the effectiveness of the improved action approach in removing
finite-spacing artifacts.  The use of anisotropic lattices in which the
temporal spacing is much smaller than that in the spatial directions
allows much more efficient glueball mass measurements.}

Glueballs and hybrid mesons are presently of great interest
theoretically and experimentally.  The lattice formulation of
QCD provides an ideal setting in which to carry out theoretical
studies of such systems from first principles using sophisticated
numerical simulations.  In order to extract the physical properties
of glueballs and hybrid mesons from such simulations, systematic
errors from the finite lattice spacing $a$ must be removed or
made acceptably small.  There are two approaches to accomplishing
this:  (1) using finer grids or (2) using improved actions on coarse
grids.  The first approach is much simpler and has been used in
almost all previous glueball and hybrid meson studies.  However,
this approach requires vast computational power.  As the grid is
made finer, many more lattice sites are needed to maintain the
physical volume of the system.  The simulation costs rise
typically as $1/a^6$ as $a$ is decreased.  Because of this,
lattice studies of glueballs have in the past been dominated by
large collaborations using some of the world's fastest
supercomputers.  

Here, we show that the second approach,
the use of improved actions, can be used to study glueballs much
more efficiently.  Improved actions have smaller lattice spacing
errors, and hence, permit the use of much coarser lattices
which can be simulated using modern computer workstations.
The key to the success of the improvement program is the
reliable determination of the couplings of the interactions
terms in the action.  Much effort over the past decade has been
directed towards this problem.  Recently, two competing
methods have emerged, one which uses block-renormalization
group transformations, another which advocates a judicious
combination of mean field theory and perturbation theory.
In this work, we use the latter approach.

A novel feature of our calculations is the use of anisotropic
lattices in which the temporal spacing $a_t$ is much smaller than
the spacing $a_s$ in the spatial directions.  
This allows much more efficient glueball mass measurements by
exploiting the enhanced signal-to-noise of the glueball correlation
functions at smaller temporal separations.  Mean-field link
renormalization\cite{TI} is crucial for maintaining the proper
renormalized anisotropy $a_t/a_s$.

\begin{figure}[t]
\psfig{figure=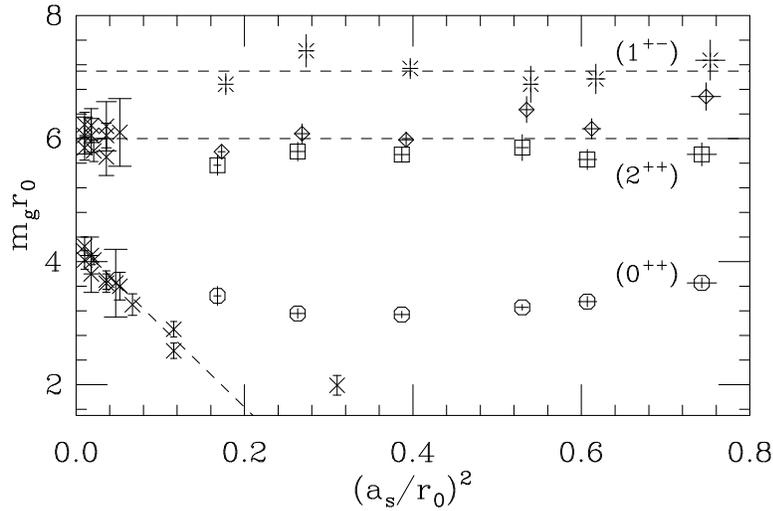,width=4.0in}
\caption{Low-lying glueball masses $m_g$ against
lattice spacing $a_s$ in terms of the hadronic scale
$r_0\approx 1/2$ fm defined by $[r^2 dV(r)/dr]_{r=r_0}=1.65$
where $V(r)$ is the static quark potential.
The $\times$ denote results for the scalar and tensor glueballs
using the simple Wilson action\protect\cite{WG}.  Results
using the improved action for the $A_1^{++}$, $E^{++}$, $T_2^{++}$,
and $T_1^{+-}$ representations are indicated by
$\circ$, $\Box$, $\Diamond$, and $\ast$, respectively.
\label{fig:scaling}}
\end{figure}

Our results are shown in Fig.~\ref{fig:scaling}.  The
scalar glueball mass from the improved action exhibits dramatically
reduced cutoff contamination compared to the Wilson action.
Finite-$a_s$ errors are seen to be small for the tensor and
pseudo-vector glueballs, although differences between
the $E^{++}$ and $T_2^{++}$ representations indicate small
violations of rotational invariance, especially for large $a_s$.
These results clearly show that glueballs can be studied without
the use of supercomputers, provided that simulations are carried
out using improved actions on anisotropic lattices.


\section*{References}
\begin{thebibliography}{99}

\bibitem{WG} J.~Sexton, {\em et al.},
  \Journal{\PRL}{75}{4563}{1995}; P.~De Forcrand {\em et al.},
  \Journal{\PLB}{152}{107}{1985}; C.~Michael and M.~Teper, 
  \Journal{\NPB}{314}{347}{1989}; UKQCD Collaboration,
  \Journal{\PLB}{309}{378}{1993}.
\bibitem{TI} G.P.~Lepage and P.B.~Mackenzie, 
  \Journal{\PRD}{48}{2250}{1993}.
\end{thebibliography}
\end{document}